# Controlled Coupling and Occupation of Silicon Atomic Quantum Dots

M. Baseer Haider[1,2]*, Jason L Pitters[2]*, Gino A. DiLabio[2], Lucian Livadaru[1,2], Josh Y Mutus[1,2] and Robert A. Wolkow[1,2]**

*Authors contributed equally

**Corresponding author

[1]Department of Physics, University of Alberta, 11322 - 89 Ave., Edmonton, Alberta T6G 2G7, Canada

[2]National Institute for Nanotechnology, National Research Council of Canada, 11421 Saskatchewan Drive, Edmonton, Alberta T6G 2M9, Canada

**It is discovered that the zero-dimensional character of the silicon atom dangling bond (DB) state allows controlled formation and occupation of a new form of quantum dot assemblies. Whereas on highly doped n-type substrates isolated DBs are negatively charged, it is found that Coulomb repulsion causes DBs separated by less than ~2 nm to experience reduced localized charge. The unoccupied states so created allow a previously unobserved electron tunnel-coupling of DBs, evidenced by a pronounced change in the time-averaged view recorded by scanning tunneling microscopy. Direct control over net electron occupation and tunnel-coupling of multi-DB ensembles through separation controlled is demonstrated. Through electrostatic control, it is shown that a pair of tunnel-coupled DBs can be switched**

**from a symmetric bi-stable state to one exhibiting an asymmetric electron occupation. Similarly, the setting of an antipodal state in a square assembly of four DBs is achieved, demonstrating at room temperature the essential building block of a quantum cellular automata device.**

There has been great interest of late in discrete solid-state structures relevant to nano-electronic and quantum computing applications. Important candidate entities are semiconductor quantum dots (QD) and impurity state tunneling systems.[1,2,3,4,5,6] In the area of coupled quantum dots great progress has been made in gaining quantum state control, in coherent manipulation and in quantum state storage.[7,8,9,10,11] Studies of quantum cellular automata (QCA) [12,13,14] involve "cells" composed of multiple QDs, each with dimensions on the order of tens of nm.[15,16,17,18,19] In the QCA paradigm multi-cellular structures transmit binary information and perform computations at extremely low energy cost. To date, electrostatic QCA embodiments have required temperatures in the milli-Kelvin range and local electrostatic tuning to achieve the appropriate electron filling and to nullify the effects of stray charges. In general, zero-dimensional systems offer a rich territory for exploratory device concepts.

Here we demonstrate that single silicon dangling bonds (DB) on an otherwise hydrogen-terminated silicon crystal surface can serve as quantum dots. Unlike the delocalized valence and conduction band states in a silicon crystal, DB states exist within the silicon band gap. DBs are therefore substantially decoupled from the bulk and retain atom-like character. Here, the distance-dependent coupling of two or more DBs via an electron tunneling interaction is described. Working under conditions where individual DBs are ordinarily negatively charged, coupled DBs are found to exhibit a "self biasing" behavior – Coulombic repulsion acts to reduce electron filling and enable tunneling betweens DBs. The fabrication and room temperature electrostatic setting of the state of a 1×1 nm assembly composed of four coupled silicon DBs is demonstrated. This assembly is reminiscent of a single QCA cell.

Scanning tunneling microscopy (STM) images of H-terminated silicon surfaces are shown in Fig. 1. Each surface silicon atom at the (100) surface shares in a surface-parallel Si-Si dimer bond and 2 bonds to Si substrate atoms, and as well is bonded to a single capping H atom, as shown in the Fig. 1 inset. The diagonal bar-like features in the images are rows of silicon dimers. Fig. 1a shows a distribution of dangling bonds. DBs exist at surface silicon atoms that do not have a capping H atom. The surface imaged in Fig. 1a has a low n-type doping concentration and as a result the DBs tend to be neutral, corresponding to one electron in the DB state, as discussed previously.[20]

The high Fermi level associated with the highly n-type doped H-terminated silicon shown in Fig. 1b, causes the DB gap states to be occupied by one extra electron. The DBs on this surface contain a total of 2 electrons, rendering them negatively charged. The charged DBs bend bands upward, thereby locally inhibiting electron injection from the STM tip into the conduction band (CB) and causing a dark halo to surround DBs in unoccupied state images, as seen in Fig. 1b.[21] DBs show the dark halo regardless of whether they are created by single H atom removal with the STM tip,[22] or are naturally occurring, as a result of incomplete H-termination. The imaging characteristics of dangling bond states are consistent with previous STM studies on surface states.[21,23]

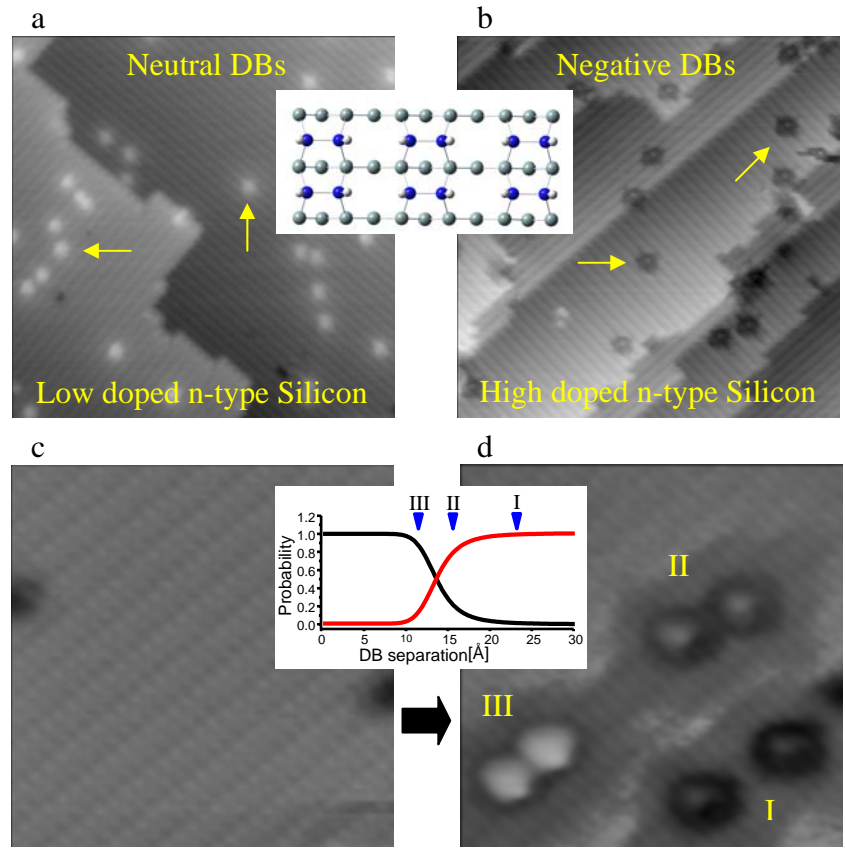

Figure 1. Comparison of DB images on hydrogen terminated Si(100) based on bulk doping densities (a, b) and separation distance (c, d). The upper inset shows the surface structure – silicon dimers are shown in blue, subsurface silicon atoms in grey and hydrogen atoms in white. a) Low doped n-type silicon (~$10^{16}$ cm$^{-3}$). 35x35 nm, 2 V, 0.1 nA. Dangling bonds appear as bright spots (two are indicated with arrows). b) High doped n-type silicon (~$10^{19}$ cm$^{-3}$) 35x35 nm, 2.2 V, 0.1 nA. Dangling bonds appear as dark depressions with a central spot (two are indicated with arrows). c) 9x9 nm, 2 V, 0.2 nA. The silicon surface appears equivalent prior to hydrogen desorption. d) 9x9 nm, 2 V, 0.2 nA. Three groups of dangling bonds are prepared. (I) A non-coupled DB pair at 2.32 nm. (II) A coupled DB pair at 1.56 nm. (III) A coupled DB pair at 1.15 nm. The lower inset indicates the charging probabilities of a DB pair based on separation distance. The black and red curves represent one and two electron occupation respectively. Positions I, II, and III correspond to the related DB pairs in part (d).

Figs. 1c and 1d explores the effect of distance between DBs. It is evident that closely spaced DBs take on a "brighter" appearance than relatively isolated DBs. It is also seen that beyond some threshold, further reduced separation leads to further enhanced brightening. This pronounced effect has never before been reported. In Fig. 1d an STM procedure for removing single H atoms was used to make three pairs of DBs with varying separations. Both DBs of pair I image as individual, i.e. uncoupled, negatively charged DBs. However, pairs II and III display a distance-dependant brightening. A DB within approximately 15 Å of another DB will exhibit this coupling, which is slightly tuned by the particular angular arrangement of a DB pair with respect to the surface lattice. The extreme cases of two DBs on one silicon dimer[24] or on adjacent dimers in a row[25] are well known and qualitatively different than the weaker coupling considered here.

Figs. 2a and 2b demonstrate that an isolated negative DB (labeled DB2) will enter into a coupling arrangement when offered a sufficiently close partner (DB3). The new DB in Fig. 2b was formed by selective single H atom removal. Once paired, DB2 loses the halo characteristic of an uncoupled, charged DB. Figs. 2c and 2d demonstrate the elimination of the coupling effect by capping one of a pair of DBs (DB6) with a hydrogen atom. The remaining DB (DB5) becomes effectively isolated and negatively charged.

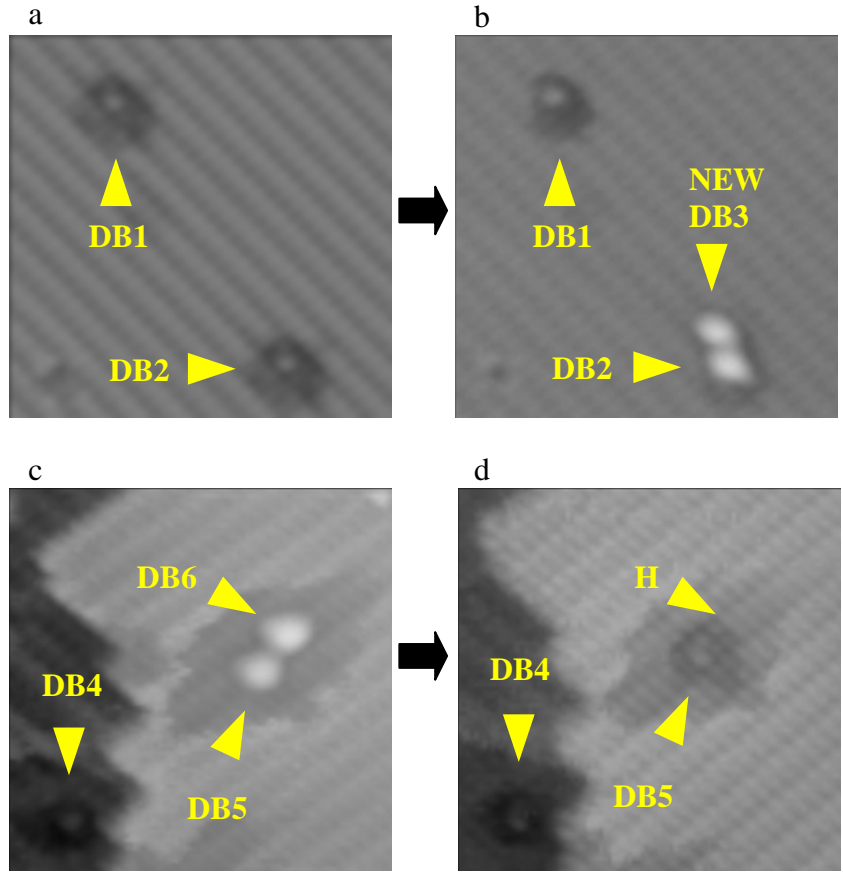

Figure 2: Coupling and uncoupling of dangling bonds on hydrogen terminated Si(100). a)10x10nm, 2V, 0.2nA. Two negative dangling bonds are indicated DB1 and DB2. b) 10x10nm, 2V, 0.2nA. A new DB is created by STM induced H desorption in close proximity to DB2. DB2 and DB3 are now coupled. c) 9x9nm, 2V, 0.2nA. DB4 is isolated and negative. DB5 and DB6 are coupled. d) 9x9nm, 2V, 0.2nA. A hydrogen atom (H) caps DB6. DB5 is now isolated and has become negative.

It is apparent that the coupling between two sufficiently close DBs is mediated by the lattice, as through-space covalent bonding is negligible beyond DB separations of ~4 Å. Calculations at zero temperature were performed on a silicon cluster model containing two P (dopant) atoms using hybrid density function theory.[26,27,28] The calculations reveal that the DB electrons are partially localized in the vacuum region above the surface, but have greater spatial breadth in the Si medium. The results also show that as two negatively charged DBs are stepped closer together, the average energy of the DB states increases. Ultimately, the Coulombic repulsion between the DBs is large enough to raise the energy of the DB states such that it is energetically more favourable to exclude one of the extra electrons from one of the DBs. Additional information is provided in the Supporting Information Section.

To further explore the DB coupling, we modeled the DBs using a harmonic oscillator (HO) potential, as was previously done for coupled quantum dots.[29] An effective two-dimensional HO, modified to allow electron escape into the silicon conduction band, is used to describe the electron in a DB In this simple model, the isolated DB wavefunction has the form of a Gaussian-type function. (see Supporting Information for additional details). The extra electron will tunnel between neighboring DBs, provided that their separation is small enough, and that Coulomb repulsion prevents full occupancy of the two DBs.

In Fig. 3, we show how our model provides a qualitative description of the behavior of two isolated DBs. The potential wells in Fig. 3 plateau asymptotically at the CB energy.

Fig. 3a depicts two potential wells, sufficiently separated laterally to be uncoupled. The horizontal line crossing each well represents the bound DB level, which is capable of holding up to two electrons each. Fig. 3b represents 2 DBs laterally separated by less than ~15Å. As a result of Coulombic repulsion, the occupation state with 2 electrons per DB is destabilized (shifted upward in energy). The two DB states are shown to be more stable if one electron is excluded from one of the DBs. The excluded electron goes into a bulk energy level. This is a result of both an on-site (repulsive) pairing energy – modeled as the "Hubbard $U$" term[30] - but also and more substantially increased Coulombic repulsion when two electrons are so localized. The effect of tunneling on electron energy levels is also included via the tunneling splitting energy $t$.[31] This exclusion of an electron from a DB pair is the manifestation of a "self-biasing" effect and is crucial to the coupling of these closely spaced atomic quantum dots.

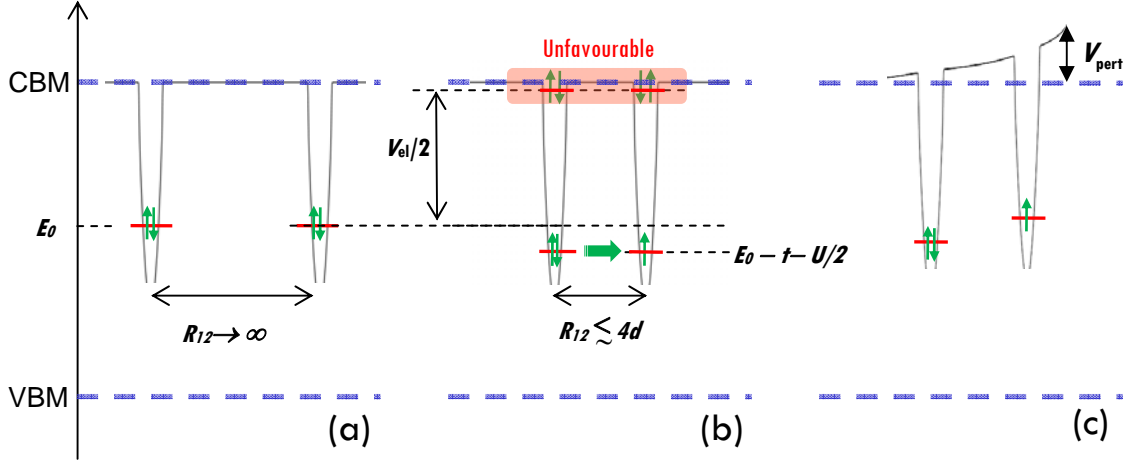

**Figure 3:** Paired DBs are illustrated as variably spaced potential wells. (a) shows isolated DBs, each negatively charged (a neutral DB has one electron occupation; a negative DB is occupied by two electrons at a ground state energy $E_0$). In (b) we show the result of Coulombic repulsive interaction $V_{el}$ and how one electron is excluded from a pair of coupled DBs, resulting in a net charge of one electron. The occupation state in which each DB has one extra electron, shaded in red, becomes increasingly unfavourable with decreasing distance. The repulsive Hubbard on-site pairing energy $U$ and the tunnel splitting energy $t$ are also indicated, and $d$ is the nearest dimer-dimer distance. (c) indicates a perturbed double-well potential for two tunnel coupled DBs in the vicinity of a third more distant negatively charged DB. Here, we assume a small surface coverage of DBs, so that band bending is negligible at the surface, except for in the immediate vicinity of a charged DB. CBM and VBM are the bulk conduction band minimum and the valence band maximum, respectively.

Two conditions allow electron tunneling between coupled DBs. When close enough, the barrier separating the DBs is sufficiently narrow to permit substantial tunnel exchanges between the two centers. But crucially, the exclusion of one electron from the paired DBs provides a partially empty state and therefore a destination for a tunneling electron. Our computations lead to an explanation for the relatively bright appearance of coupled DBs: Upon expulsion of one of the 2 extra electrons upward band bending is reduced, enabling relatively easy injection of electrons from the STM tip to the CB and a brighter appearance. Though the charge at coupled DBs is reduced, resulting in an appearance more like that of neutral DBs (Fig. 1a), the pair-encircling dark halo due to an extra electron, distinguishes coupled DBs from neutral DBs.

The increased brightness of a DB pair with decreasing DB separation is a manifestation of the crossover between charging states –2 to –1. The injected STM current increases because of decreasing local band bending. A more detailed statistical treatment of occupation of coupled DBs, presented in the Supporting Information, reveals that average electron occupation of coupled DBs shifts from just under 2 electrons for DBs spaced by approximately 23 Å to 1 electron as DBs become closer than 10 Å, accounting for the increasingly bright appearance of DB pairs with decreasing separation, as seen in Fig. 1d. A graph showing the electronic occupation as a function of DB separation is given in Fig. 1. It is clear that the images in Figs. 1 and 2 represent direct observations of tunnel-coupling between quantum dots.

We note that the tunnel-coupled pairs of Si DBs resemble charge qubits that are being considered for quantum computing architectures. However, it remains to be demonstrated that these tunnel-coupled pairs have the required coherence characteristics of qubits. This is the subject of an on-going investigation.

We now examine assemblies of more than two quantum dots. Fig. 4a shows 3 coupled DBs. The question of how many extra electrons are contained by such a structure is in part answered by our experimental observations: The encircling dark halo indicates that the assembly is negatively charged. The brightness variation among the DBs indicates that DB1 and DB3 are more negatively charged, on average, than DB2. This can only be caused by the presence of two extra electrons in the assembly. This is supported by calculations which show that Coulombic repulsion is too great for 3 electrons to be bound on the structure shown in Fig. 4a

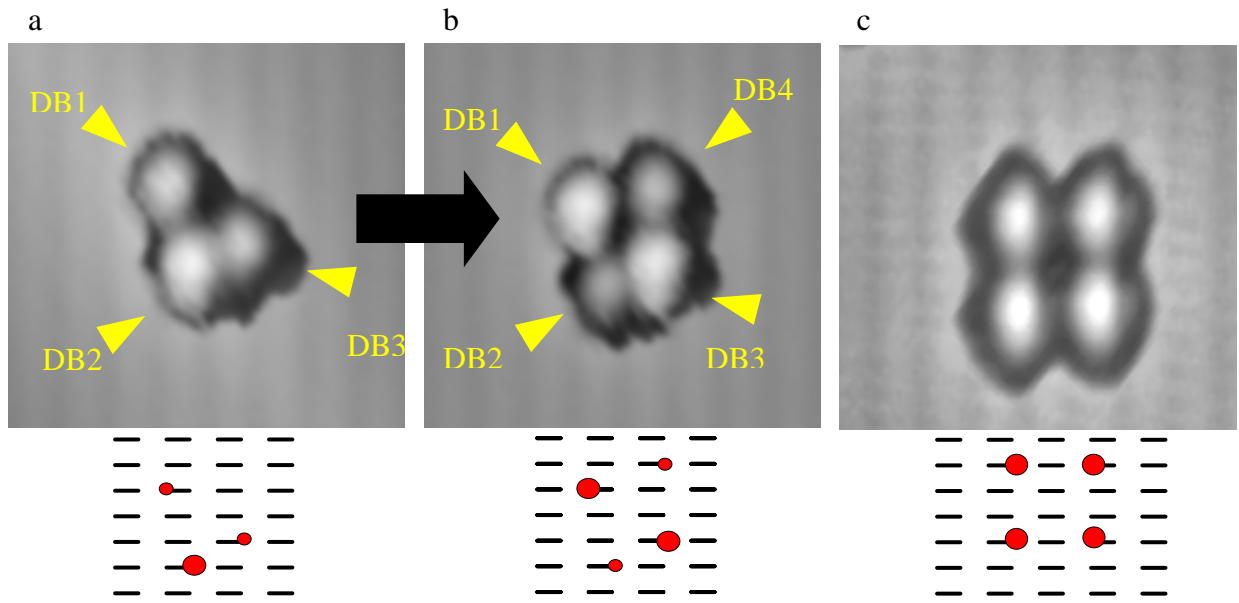

Figure 4. Coupling in asymmetric and symmetric DB arrangements. a) 6x6 nm, 2V, 0.1nA. Shows 3 coupled DBs. Clear inequivalencies exist between the 3 DBs. DB2 is brightest while DB1 and DB3 appear darker. b) 6x6nm, 2V, 0.1nA. Shows the same area after a fourth coupled DB was created by removal of a single H atom. DB2 has become one of the less bright features while DB1 and DB3 are now brightest. The average height difference between the bright and less bright DBs is ~0.4 and ~0.7Å for (a) and (b) respectively. c) 6x6 nm, 2V, 0.08 nA. A separate experiment showing a rectangular symmetric group of 4 DBs created by the removal of four single H atoms. The DB heights are equal within 0.1 Å. Grids are shown below the figures to represent the DB positions on the silicon surface. The dashes represent silicon dimers and red circles represent coupled silicon DBs. The size of the red circle illustrates the intensity of the DB as it appears in the above STM images.

In Fig. 4b a fourth coupled DB (DB4) is created by STM tip-induced removal of a single H atom. The schematic in Fig. 4 shows that the four DB structure deviates from a regular rectangular arrangement. The most widely separated DBs in the group of four DBs (DB2 and DB4) are darkest in appearance. Note that DB2 was initially brighter than DBs 1 and 3 in Fig. 4a, then became darker than DBs 1 and 3 in Fig. 4b. It is apparent that the extra electrons in this structure are predominantly located at the most distant DBs (DBs 2 and 4). This is consistent with the expectation that the greatest charge separation corresponds to the lowest energy configuration. Computation reveals that a net charge of -3 on this 4 DB assembly is prohibitive due to excessive Coulombic repulsion. The experimental observations are consistent with a charge state of -1 because a single electron would be shared between the two most highly coupled DBs. This would result in adjacent (rather than opposite DBs) appearing darker. A charge state of -2 fits the experiment and computational results. Fig. 4c shows a rectangular arrangement of 4 coupled DBs. In this case the DBs appear indistinguishable indicating that the 2 extra electrons are equally shared between all 4 DBs. This contrasts the case shown in Fig. 4b where the irregular structure determines the asymmetric distribution of extra electrons.

The structure shown in Fig. 4c is reminiscent of the central building block in the QCA scheme proposed by Lent and co-workers in 1993.[12] However, to serve as a QCA cell it must be possible to induce an asymmetry in the electronic structure of the cell, which can be mapped to binary states "0" and "1". This could be achieved on silicon through the use of an applied electrostatic field. In order to explore this possibility, we first return to the simpler case of a 2 DB cell to consider the effect of a point electrostatic perturbation.

Fig. 5a shows a pair of coupled DBs that share one extra electron. The DBs have only a small asymmetry. A third, negatively charged DB (DB3) was created at a distance where it cannot tunnel-couple with DBs 1 and 2 (Fig. 5b). However, the electrostatic field emanating from DB3 exerts a repulsive effect which causes the extra electron in the DB pair to favor occupation on DB1. This effect is clear in the line profile shown in Fig. 5c: An electrostatic perturbation has caused DB2 to become appreciably brighter, and therefore less negative, than DB1. This electrostatic perturbation effect is an example of the type of gating required in a QCA device.[14,15] It is a breaking of symmetry leading to a dominance of one occupation over another. Note that the "self-biasing" effect – the determination of occupation through control of DB separation - results in the two DB entity having only one extra electron, removing the need for a "filling" gate to tune occupation. Fig. 3b illustrates this effect.

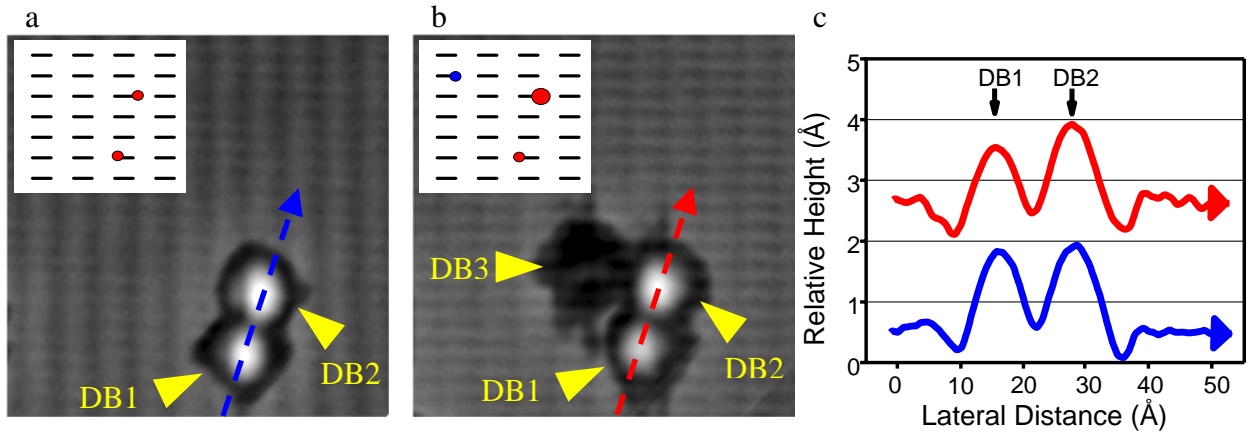

Figure 5: Demonstration of an electrostatic perturbation on coupled DBs. a) 8x8nm 2V, 0.08nA. Two coupled DBs. The height profile of the DB pair is indicated in blue and is shown in part c. Only a small asymmetry is evident in the cross sectional view. b) 8x8nm 2V, 0.08nA. The same area after the addition of a third DB. DB3 is negatively charged and not coupled to DB1 and DB2. The height profile of the DB pair is indicated in red and is shown part c. DB2 now appears much brighter as a result of its proximity to the negatively charged DB3. The insets show grids to represent the DB positions on the silicon surface. The dashes represent silicon dimers, red circles represent coupled silicon DBs and the blue circle represents an uncoupled, negatively charged silicon DB. The size of the red circle illustrates the intensity of the DB as it appears in the above STM images. c) Comparison of relative height profiles for the DB pair prior to (blue) and after (red) the addition of a perturbing DB (DB3). The line profiles are offset for clarity.

Returning to the properties of a rectangular four dot cell, Fig. 6 shows 4 coupled DBs and 2 diagonally placed perturbing DBs. The schematic shows the positions of all 6 DBs. Consistent with the expectation that a negative gating effect destabilizes electron occupation, it is seen that the two DBs within the cell that are nearest the perturbing DBs appear brighter, meaning less negative charge is localized there  The image shows a controlled electrostatic breaking of symmetry and setting of an antipodal state.  This represents an atomic scale embodiment and simplest functional demonstration of a QCA cell.

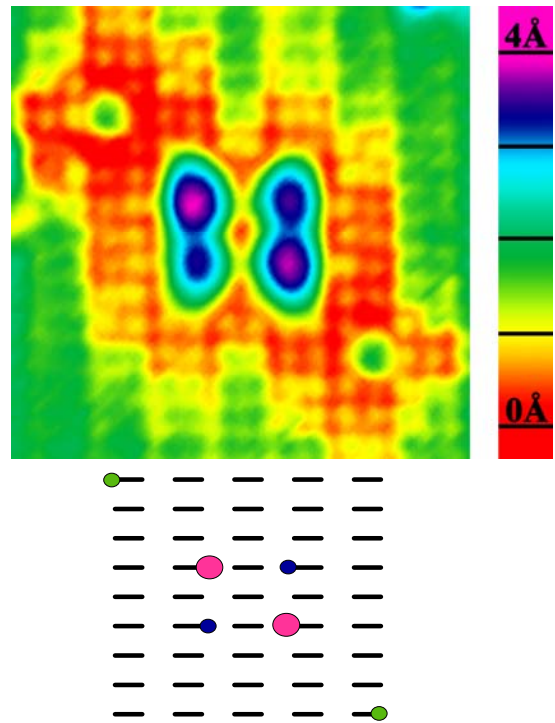

Figure 6. A colour mapped STM image (6x6nm, 2.5V, 0.11nA) of a rectangular four coupled DB entity with an additional two electrostatic perturbing DBs diagonally placed. The 2 coupled DBs nearest the negative perturbing DB are relatively high in appearance as a result of unfavored electron occupation at those sites. The average height difference between the violet (higher) and blue colored (lower) DBs is ~0.4Å. A grid is shown below the figure to represent the DB positions on the silicon surface. The dashes represent silicon dimers, purple and blue circles represent coupled silicon DBs and green circles represent uncoupled, negatively charged silicon DBs. The size of the circles illustrates the intensity of the DB as it appears in the above STM image.

Within this scheme, considerable latitude exists for control over electron occupation. Slightly more widely spaced configurations will reduce Coulombic repulsion to allow, and lead automatically to, a greater net charge. Similarly, closer spaced structures will naturally exhibit reduced charging. This "self biasing" is most desirable as it removes the need for multiple "filling" gates. A related result is notable: Fixed charges beyond ~20 Å are calculated to be insignificant perturbations on the state and filling level of a Si DB-based QD assembly. This is in contrast to the acute sensitivity of some quantum dots to relatively distant unintended charges. [32] The robustness of the atomic system described here results from the relatively great energy level spacing of bound states. For this same reason, a QCA cell embodied of Si DBs is capable of performing at room temperature rather than requiring cryogenic conditions. It is anticipated that coupled atomic quantum dots composed of other materials will also exhibit similar properties.

Future efforts are focused on the fabrication and characterization of multi-cell structures and variable rather than fixed gating. Control over these assemblies may be achieved through directed molecular attachment, by adjusting the local doping level or by using electrodes to vary the local electrostatic potential.

## ACKNOWLEDGEMENTS

This research was supported by the Canadian Institute for Advanced Research, the Natural Sciences and Engineering Research Council of Canada, the Informatics Circle of Research Excellence and Alberta Ingenuity. GAD thanks the Centre for Excellence in Integrated Nanotools (University of Alberta) for access to computational resources. The authors are grateful to Martin Cloutier and Dr. Radovan Urban for technical assistance.

# Supporting Information

## Theoretical Model

**Details of the simplified model for electron confinement and tunneling in dangling bond systems**

Density functional theory modeling was performed using the B3LYP [1, 2] functional, as implemented in the Gaussian-03 [3] package, with 3-21G basis sets. The model used for the calculations consisted of pyramidal cluster of 396 silicon atoms with a 2x1 surface reconstruction and had three rows of seven dimers. The cluster contained two P dopant atoms. Two P atoms were required in order to provide the cluster with the two excess electrons needed to charge two of the surface dangling bonds (DBs). One DB was stepped along one side of a dimer row toward a second DB of fixed position. A third DB, placed near one corner of the 2x1 face of the cluster, acted as an electron accepting site for the electron that is excluded from the closely-spaced DB pair when the Coulombic repulsion there becomes large.

While models of this type are too small to reproduce the band gap in H-terminated silicon, they are sufficiently large to predict some key features of the surface DB states, in particular a neutral DB state resides 0.35 eV above the valence band edge.

B3LYP calculations on a smaller cluster were used to provide a picture of the one-electron state of the charged DB. The physical extent on the DB state is clearly bounded by the row structure of the H-Si(100)-2x1 surface (Figure S1).

In order to explore DB coupling, we use a harmonic oscillator potential, as previously proposed for coupled quantum dots [4]. Our confinement potential for an isolated DB allows for electron escape from the DB orbital into the bulk conduction band. Therefore, outside the harmonic well region $r < R_h$ the potential curve levels off to the bulk conduction band minimum (CBM), which is thus taken as the zero value for the confinement potential at $r \to \infty$.

Thus, a finite/truncated confinement potential is chosen as

$$V_{DB}(|\mathbf{r}|) = \left(\frac{m\omega^2}{2} r^2 - V_0\right) f_{tr}(r), \qquad (1)$$

where $\omega$ is the classical oscillation frequency of the electron in the well, $V_0$ is the potential depth (measured from the CBM), and $f_{tr}$ is a truncation function, which ensures the finite range of the harmonic potential. We choose a simple form for

$$f_{tr}(r) = [1 - \tanh((r - R_b)/w)]/2, \qquad (2)$$

where the parameters $R_b$ and $w$ determine the location and the width of the truncation region, respectively.

In this simplified model, the isolated DB wavefunction has the form of a simple Gaussian-type function, $\psi_0(\mathbf{r}) = A\exp(-\alpha r^2)$, and a ground state energy of $E_0 = \hbar\omega$ (with respect to the potential bottom), where $\alpha = m\omega/2\hbar$, and $A$ is a normalization constant. The parameter $\omega$ is set such that the spatial extent of the simple Gaussian DB wavefunction reflects that obtained from our DFT modeling (ca. 3.8 Å). The binding energy of an electron in the HO well is then $E_b = E_{CB} - E_0$. The well depth is fitted to recover the binding energy of the HO as calculated by DFT methods.

In Figure S2 we show the projection of the confinement potential along a line parallel to the surface which passes through the center of the DB, and we label the features of the potential. A double well potential for describing a tunnel-coupled DB pair was simply built by combining the single well potentials of the two DBs according to

$$V_{12}(\mathbf{r}) = \frac{1}{2}V_{DB1}(|\mathbf{r}-\mathbf{r}_1|)\left[1 - \tanh\frac{|\mathbf{r}-\mathbf{r}_1|-R_h}{w}\right] + \frac{1}{2}V_{DB2}(|\mathbf{r}-\mathbf{r}_2|)\left[1 - \tanh\frac{|\mathbf{r}-\mathbf{r}_2|-R_h}{w}\right], \qquad (3)$$

which has the effect of $V_{12}(\mathbf{r})$ being approximately equal to either $V_{DB1}$ or $V_{DB2}$, depending on which DB is closer to the point $\mathbf{r}$. Here, $V_{DBi}$, $i=1,2$ are the isolated DB potentials of the form in Eq. (1), with the truncation function identically 1.

For a double well system, the tunneling coefficient calculated within the WKB approximation [5] is

$$D = \exp\left\{-\frac{2}{\hbar}\int_{-a}^{a}\sqrt{2m(V_{12}(r)-E)}dr\right\}, \qquad (4)$$

where $\pm a$ are the classical turning points of the potential barrier. Accordingly, the tunneling splitting interaction is calculated as

$$t = \frac{\hbar\omega}{\pi}\exp\left\{-\frac{1}{\hbar}\int_{-a}^{a}\sqrt{2m(V_{12}(r)-E)}dr\right\}. \qquad (5)$$

**Charging states of dangling bond assemblies**

Consider a cell composed of $n$ DBs on the H-Si (100) surface. In order to calculate the charging probabilities of this cell with a number of $i$ electrons, we have to consider the effects of the DB system being in contact with a reservoir (the bulk crystal), with a temperature $T$ and a chemical potential $E_F$ (i.e. the grand canonical ensemble). Below we calculate the probability of occurrence of a charging state with a given number of extra electrons in the DB cell.

In order for a DB cell to have an integer number of extra electrons, and a well-defined polarization (as required by a QCA device), we need to be in a regime where the tunneling energies are much smaller than the Coulomb energies, $t \ll V_{el}$ [6]. Therefore below we neglect the tunneling contributions to the charging energies. At room temperature, due to statistical fluctuations, all charging states have a finite probability of occurrence. Assuming classical statistics, each charging state is characterized by a statistical weight, proportional to its Boltzmann factor. Thus, for one electron charging, the weight is

$$f_{1e} = g_{1e} \exp\left[-(E_{1e}^{tot} - E_F)/kT\right] \tag{6}$$

where $E_{1e}^{tot}$ is the total energy of the $n$-DB cell with one extra electron, $k$ is the Boltzmann constant, $T$ is the temperature, and $g_{ie}$ is the degeneracy of the charging state $i$. Assuming there are no external perturbations, $E_{1e}^{tot} = -E_b$, where all energies are measured from the CB level. Similarly, the two-electron charging weight is

$$f_{2e} = g_{2e} \exp\left[-(E_{2e}^{tot} - 2E_F)/kT\right], \tag{7}$$

with $E_{2e}^{tot} = -2E_b + V_{2e}^{tot}$ and $V_{2e}^{tot}$ being the total electrostatic interaction energy of the cell. The latter interaction is calculated simply as a point-charge interaction

$$V_{2e}^{tot} = \frac{1}{4\pi\varepsilon_0\varepsilon_{srf}} \frac{e^2}{d}, \tag{8}$$

where $e$ is the elementary charge, $d$ is the DB separation, and $\varepsilon_{srf}$ is the effective dielectric constant of the surface.

Similarly, the weight of any charging state $i$ is

$$f_{ie} = g_{ie} \exp\left[-(E_{ie}^{tot} - iE_F)/kT\right]. \tag{9}$$

In calculating the total charging energies of the cell, we neglected the tunneling energies as noted above, and only considered the lowest energy configuration for a given charging state (e.g. diagonal occupancy only for a doubly charged 4-DB cell, and no allowed configuration with the electrons on adjacent DBs).

The partition function of the $n$-DB cell is then

$$Z = f_{1e} + f_{2e} + ... + f_{ne}, \tag{10}$$

and the probabilities of each charging state $i$ are then $p_{ie} = f_{ie}/Z$, for $i=1,n$. For instance, in order to have a "definitely" doubly occupied 4-DB cell, we need to fulfill the condition $p_{1e}, p_{3e}, p_{4e} \ll p_{2e} \cong 1$.

In the lower inset of Figure 1 we show the charging probabilities for an isolated 2-DB cell as a function of the DB separation. The value of the binding energy $E_b$ for a doubly occupied DB was taken to be 0.32 eV. We assume a small surface coverage of DBs, so that band bending is negligible at the surface, except in the vicinity of the DB cell. The probability of charging state 0, $p_{0e}$ was calculated to be four orders of magnitude smaller than $p_{1e}$, and therefore it was neglected. We compare DB images on hydrogen terminated highly doped n-type Si(100) taken for various separation distances appearing in Figure 1d: (I) 2.32nm, (II) 1.56nm, and (III) 1.15nm, which are also marked by triangles in the lower inset of Figure 1. The effective dielectric constant for interaction between 2 charged DBs is around $\varepsilon_{srf} \cong 4$. This is lower than the classically derived value of 6.5 [7], but not surprising since the DB orbitals are partially localized in vacuum and are elevated with respect to the surrounding Si surface atoms. This value of $\varepsilon_{srf}$ is also roughly corroborated by fitting the location of the crossover between the charging states –1 and –2 to STM images of DB pairs at decreasing distances as in Figure 1.

Accounting for the charging probability of a DB, we can now estimate the tunneling rates between *two tunnel-coupled DBs* as the product of the frequency of attempts $\omega/2\pi$, the tunneling coefficient $D$, and the probability of having an unoccupied second DB when the first DB is occupied, equal to $p_{1e} = 1 - p_{2e}$.

$$k_{tun} = \frac{\omega}{2\pi} D p_{1e}, \qquad (11)$$

where $p_{1e}$ is the probability of having 1 extra electrons in a 2-DB cell (black curve in the graph in Figure 1). The results for tunneling rates are shown in Figure S3.

**Reference**

S1. A. D. Becke, *J. Chem. Phys.* **98,** 5648 (1993).
S2. C. Lee, W. Yang, R. G. Parr, *Phys. Rev. B* **37,** 785 (1988).
S3. M. J. Frisch *et al.*, Gaussian 03, Revision C.02, Gaussian, Inc., Wallingford, Connecticut, 2004.
S4. G. Burkard, D. Loss, D. P. Di Vincenzo, *Phys. Rev. B* **59**, 2070 (1999).
S5. J. H. Weiner, *J. Chem. Phys.* **69**, 4743 (1978).
S6. C. S. Lent, P. D. Tougaw, W. Porod, G. H. Bernstein, *Nanotechnology* **4**, 49 (1993).
S7. Ph. Ebert, *Surf. Sci. Rep.* **33**, 121 (1999).

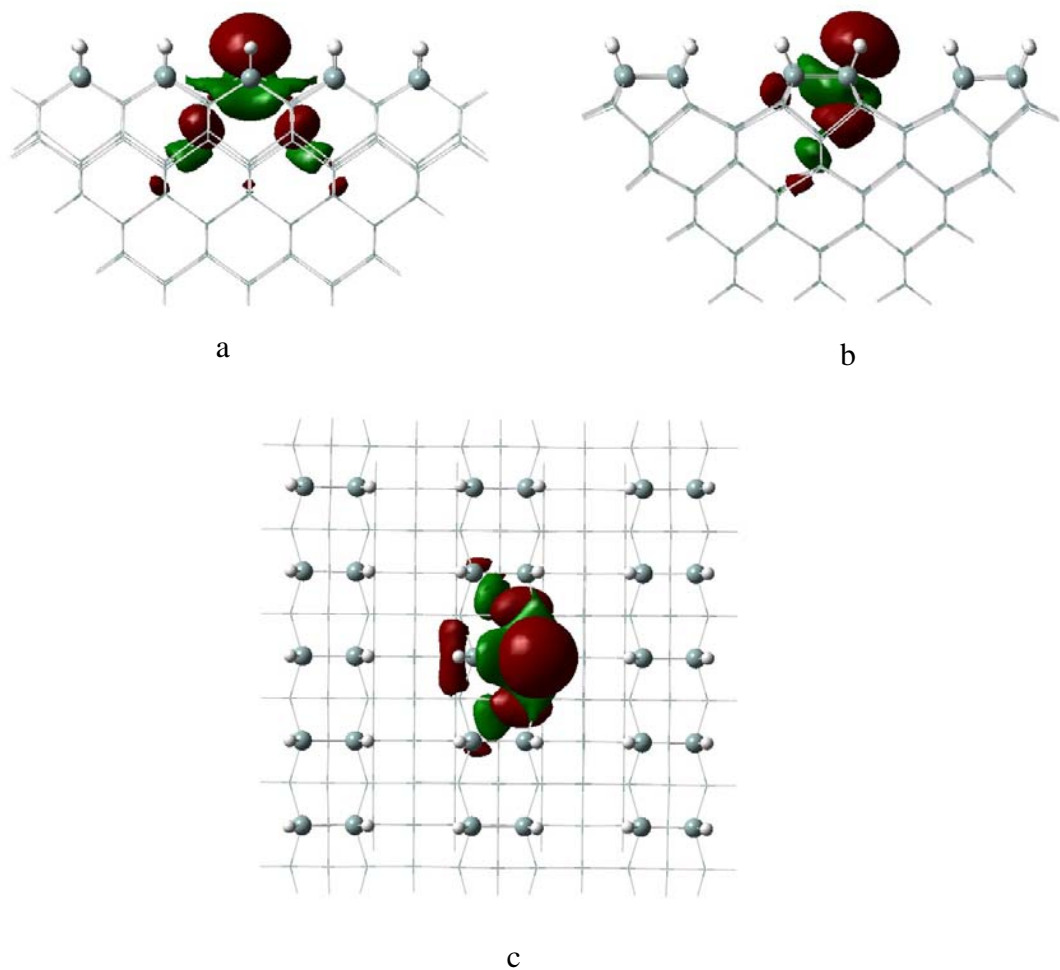

**Figure S1:** Representations of an isolated charged dangling bond state on an otherwise H-terminated Si(100)-2x1 surface obtained from B3LYP calculations. In all images, to top layer of Si and H atoms are grey and white balls, repectively, and the remainder of the cluster is represented in a stick representation. The relative phases of the dangling bond orbital are represented by the red and green rendering of the isosurface. (a) Side view from the perspective of looking along the surface dimer rows. Confinement of the DB state to a single dimer row by the vacuum surrounding that dimer row is clearly evident. (b) Side view from the perspective of looking along a direction perpendicular to the surface dimer rows. The DB state is less confined along the dimer row direction. (c) Top-down view on the DB state. The dimer rows run from top to bottom. Confinement of the DB state by the dimer row (left-to-right confinement) is evident.

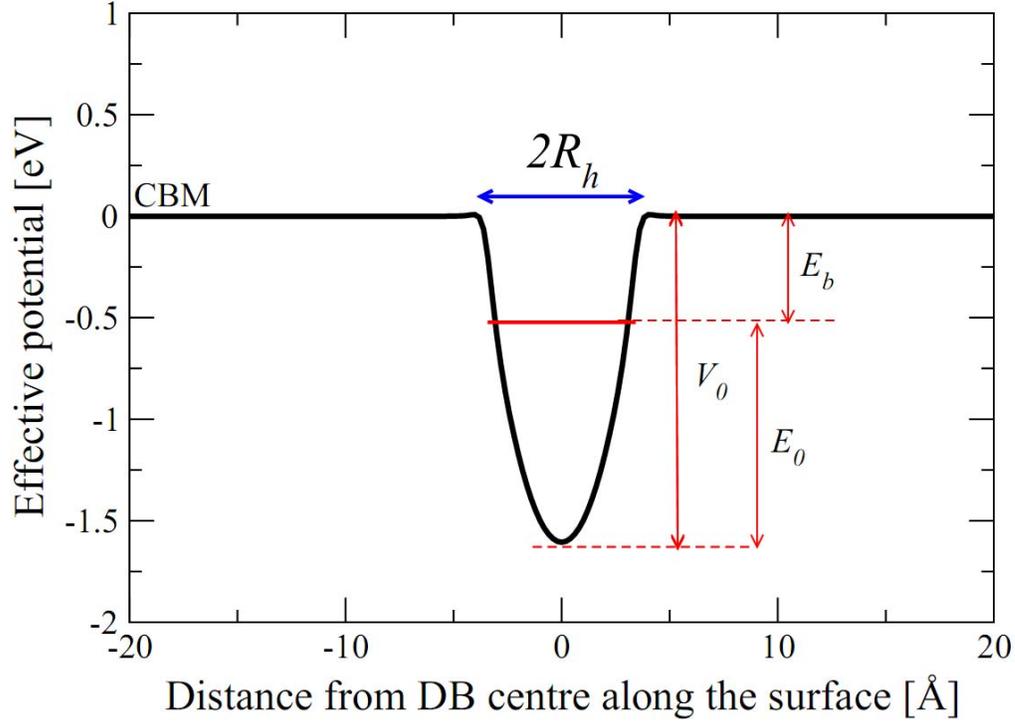

**Figure S2:** Details of the single electron confinement potential in a DB used in this study. We indicate by arrows the following: the binding energy $E_b$, the potential depth $V_0$ and the ground state energy $E_0$ (with respect to the potential bottom). The extent of the harmonic region of the potential is also shown, and the solid red line crossing the well marks the ground state level of the DB. CBM represents the location of the conduction band minimum of the bulk crystal.

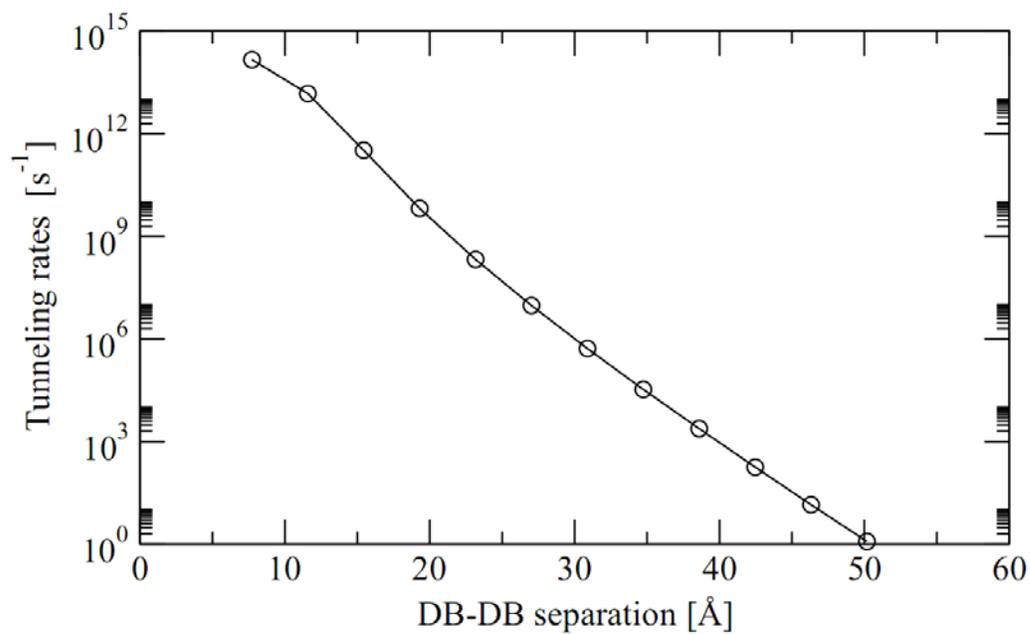

**Figure S3:** Electron tunneling rates between two neighboring DBs on hydrogen terminated highly doped n-type Si(100) as a function of DB separation. The rates were calculated according to eq. (11).